\documentclass[aip,jap,reprint,twocolumn,floatfix,nofootinbib]{revtex4-1}
\usepackage{dcolumn}
\usepackage{graphicx}
\usepackage{epsfig}
\usepackage{bm}
\usepackage{amsmath}
\usepackage{amssymb}
\usepackage{lmodern}

\newcommand{\degree}{\ensuremath{^\circ}}

\begin{document}

\title{Towards an experimental proof of superhydrophobicity enhanced
by quantum fluctuations freezing on a broadband-absorber metamaterial}
\author{Micha\"{e}l Sarrazin}
\email{michael.sarrazin@ac-besancon.fr}
\altaffiliation{Corresponding author}
\affiliation{Institut UTINAM, CNRS/INSU, UMR 6213, Universit\'{e} 
Bourgogne -- Franche-Comt\'{e}, 16 route de Gray, F-25030 Besan\c con Cedex, France}
\author{Isma\"{e}l Septembre}
\affiliation{Department of Physics, University of Namur, 61 rue de
Bruxelles, B-5000 Namur, Belgium}
\author{Anthony Hendrickx}
\affiliation{Department of Physics, University of Namur, 61 rue de
Bruxelles, B-5000 Namur, Belgium}
\author{Nicolas Reckinger}
\affiliation{Department of Physics, University of Namur, 61 rue de
Bruxelles, B-5000 Namur, Belgium}
\author{Louis Dellieu}
\affiliation{Department of Physics, University of Namur, 61 rue de
Bruxelles, B-5000 Namur, Belgium}
\author{Guillaume Fleury}
\affiliation{Univ. Bordeaux, CNRS, Bordeaux INP, LCPO, UMR 5629, F-33600, Pessac, France}
\author{Christian Seassal}
\affiliation{Universit\'{e} de Lyon, Institut des Nanotechnologies de Lyon INL-UMR 5270, CNRS, Ecole Centrale de Lyon, 36
Avenue Guy de Collongue, F-69134 Ecully Cedex, France}
\author{Radoslaw Mazurczyk}
\affiliation{Universit\'{e} de Lyon, Institut des Nanotechnologies de Lyon INL-UMR 5270, CNRS, Ecole Centrale de Lyon, 36
Avenue Guy de Collongue, F-69134 Ecully Cedex, France}
\author{S\'{e}bastien Faniel}
\affiliation{Institute of Information and Communication Technologies, Electronics and Applied Mathematics, Universit\'{e} Catholique de Louvain, Place du Levant 3, B-1348 Louvain-la-Neuve, Belgium}
\author{Sabrina Devouge}
\affiliation{Physics of Materials and Optics, Research Institute for
Materials Science and Engineering, University of Mons, Mons B-7000, Belgium}
\author{Michel Vou\'{e}}
\affiliation{Physics of Materials and Optics, Research Institute for
Materials Science and Engineering, University of Mons, Mons B-7000, Belgium}
\author{Olivier Deparis}
\email{olivier.deparis@unamur.be}
\altaffiliation{Corresponding author} 
\affiliation{Department of Physics, University of Namur, 61 rue de
Bruxelles, B-5000 Namur, Belgium}

\begin{abstract}
Previous theoretical works suggested that superhydrophobicity could be enhanced through partial inhibition of the quantum vacuum modes at the surface of a broadband-absorber metamaterial which acts in the extreme ultraviolet frequency domain. This effect would then compete with the classical Cassie-Baxter interpretation of superhydrophobicity. In this article, we first theoretically establish the expected phenomenological features related to such a kind of ``quantum'' superhydrophobicity. Then, relying on this theoretical framework, we experimentally study patterned silicon surfaces on which organosilane molecules were grafted, all the coated surfaces having similar characteristic pattern sizes but different profiles. Some of these surfaces can indeed freeze quantum photon modes while others cannot. While the latter ones allow hydrophobicity, only the former ones allow for superhydrophobicity. We believe these results lay the groundwork for further complete assessment of superhydrophobicity induced by quantum fluctuations freezing.
\end{abstract}

\maketitle

\section{Introduction}

A few years ago, it was shown, from first-principle numerical calculations,
that superhydrophobicity of nanostructured surfaces is dramatically enhanced
by tuning vacuum photon-modes via proper design of the surface corrugation,
independently of any kind of chemical functionalization \cite{s1,s2,s3}.
While nanostructures are commonly used for developing superhydrophobic
surfaces, available wetting theoretical models ignore the effect of vacuum
photon-modes alteration on van der Waals forces and thus on hydrophobicity 
\cite{w1,w2,w3,w4,w5,w6}. Quantum physics teaches us that the van der Waals
force results from the exchange of virtual photons -- i.e. quantum vacuum
fluctuations of the electromagnetic field -- between both interacting bodies 
\cite{vdW1,vdW2,vdW3,vdW4,vdW5}. Then, considering nanostructured surfaces
designed to form a thin metamaterial layer with ultra-broadband and
wide-angle absorption of electromagnetic radiation, we could preclude the
exchange of virtual photons thus inducing the collapse of the van der Waals
force \cite{s1,s2}. In this context, the study of non-wetting phenomena is a clever way to indirectly probe the van der Waals interactions and water obviously appears to be the most relevant liquid to study those effects.
In spite of this exciting possibility, experimental
investigations are still missing which would support such a quantum approach. In
this article, we address this problem from a practical point of view and
shed light on how freezing of electromagnetic quantum fluctuation allows a
kind of superhydrophobicity and could be experimentally demonstrated. In
section II, we recall the theoretical framework and provide a new practical
description and context. In section III, we report on our first experimental
attempts to observe the effect of quantum freezing on superhydrophobicity by
studying the wettability of controllable nanostructured silicon coated with
organosilane self-assembled monolayers. Our results suggest that
superhydrophobicity shows up only in those samples for which freezing of
quantum photon modes was predicted to take place. We believe these results
make a significant case in the experimental proof of superhydrophobicity
enhancement by freezing of electromagnetic quantum fluctuations.

\section{Theoretical framework}

\label{II}

\begin{figure}[!ht]
\centering
\includegraphics[width = .6\linewidth]{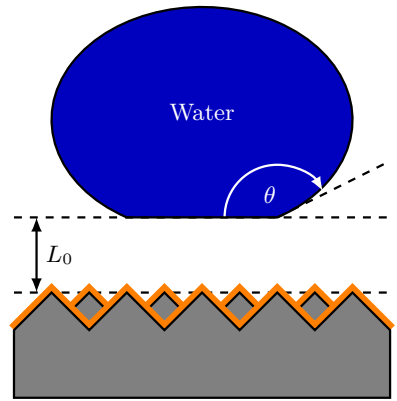}
\caption{(Color online). Sketch of the studied system. At equilibrium, the
water droplet is separated from the nanostructured surface by the distance $%
L_0$ \protect\cite{14c}. The droplet presents a contact angle $\protect%
\theta $. The nanostructured silicon (grey) is coated (orange) with organic
molecules in order to emulate a nanostructured molecular solid surface.}
\label{fig0}
\end{figure}

The main contributions to van der Waals interactions come from virtual
photon exchange in the Extreme UltraViolet (EUV) domain \cite%
{s1,s2,s3,14,14c}. As shown in our previous theoretical studies, an
ultra-broadband and wide-angle electromagnetic absorber in EUV domain should
then preclude relevant photon exchange between the absorber and a water
droplet, thus inducing van der Waals interactions between them to collapse 
\cite{s1,s2}. As the surface tension of the water droplet dominates -- over
its interaction with the broadband absorber surface -- water cannot spread
on it \footnote{Many liquids are often used to study wetting (or non-wetting) phenomena, such as diiodomethane, DMSO, ethanol, cyclohexane, or ethylene glycol. For a corrugated interface, the required condition to obtain enhanced non-wetting phenomena is that the contact angle on the flat interface on the same material is $\theta \geq 90\degree$, i.e. the flat surface is already non-wettable. This is necessary to obtain an almost flat liquid-solid interface as required for the Cassie-Baxter approach of the superhydrophobicity or for the present quantum model. From our own measurements on flat silicon grafted with organosilane molecules (the material used herein), none of the above mentioned liquids meet the expected condition. The best result was obtained for ethylene glycol with $\theta = 87.5\pm2.9\degree$ on flat grafted silicon. As a result, water is better than any other fluid for the present purpose.}. (see Fig.~\ref{fig0}). As a result, an ultra-broadband and wide-angle
absorber in the EUV domain should act as a superhydrophobic metamaterial.
Such a broadband absorber metamaterial can be designed by decorating a flat
substrate with conical nanostructures organized as a subwavelength
periodical array \cite{s1,s2}. This kind of nanostructured surface is known
to act as an optically antireflecting layer as it is equivalent to a graded
index multilayer \cite{s1,s2}. Using a numerical code based on Rigourous
Coupled Wave Analysis (RCWA) \cite{RCWA}, it is possible to compute the
scattering matrices ($S$ matrices) describing light wave interactions with
surfaces -- nanostructured or not. A relevant expression of van der Waals
forces connects these interactions to $S$ matrices at a quantum level \cite%
{s1,s2,11,10,12,13}. Between the interfaces separating two interacting
bodies by a distance $L$, the potential energy $U(L)$ related to van der
Waals forces is given by the well-known Hamaker expression \cite{14b}:

\begin{equation}
U(L)=-\frac{A_{H}}{12\pi L^{2}},  \label{Hamaker}
\end{equation}%
where $A_{H}$ is the Hamaker constant. $A_{H}$ can be derived through heavy
numerical computations of the scattering matrices describing the problem
under study. In our previous numerical studies, we considered
nanostructuring by an hexagonal array of cones as a theoretical framework.
For this kind of nanostructured surface, the Hamaker constant was computed
against the cone height $h$ \cite{s1,s2} leading to a dependence on $h$
which is well fitted by: 
\begin{equation}
A_{H}\sim A_{H,0}\frac{1}{1+h/h_{0}},  \label{Ahfit}
\end{equation}%
with $h_{0}\sim a_{0}/\pi $, where $a_{0}$ is the grating parameter -- i.e.
the spatial period of the cone array -- and $A_{H,0}$ is the Hamaker
constant of the flat surface ($h=0$). Such a convenient expression can also
be derived in the context of a simple analytical model introduced in this
article, in order to clarify the physical meaning of Eq.~\ref{Ahfit} and to
drive easily experimental investigation. For the sake of clarity, details of the following mathematical arguments are given in Appendices \ref{A} and %
\ref{B}.

Let us consider media (bodies) $1$ and $2$ occupying the half-spaces $z<0$
and $z>L$, respectively, and separated by a vacuum. At absolute zero
temperature, it can be shown that the van der Waals interaction potential
energy $U$ is given by \cite{10,11,12,13} $U=\sum_{p}\frac{1}{2}\hbar
(\omega _{p}(L)-\omega _{p}(L\rightarrow \infty ))$ where $\omega _{p}(L)$
is the eigen angular frequency -- for a given polarization -- of the p$^{th}$
vacuum photon-mode available between the two media facing each other. Using
the Cauchy's argument principle and the analytical properties of the Fresnel
coefficients of each body, the interaction energy is then given by the exact
expression \cite{10,12,13}:

\begin{eqnarray}
U(L) &=&\frac{\hbar }{2\pi }\sum_{m=s,p}\int \frac{\mathrm{d}^{2}k_{%
\mathbin{\!/\mkern-5mu/\!}}}{(2\pi )^{2}}\int_{0}^{\infty }\mathrm{d}\xi
\label{U} \\
&&\times \ln (1-R_{1}^{m}(i\xi ,\mathbf{k}_{\mathbin{\!/\mkern-5mu/\!}%
})R_{2}^{m}(i\xi ,\mathbf{k}_{\mathbin{\!/\mkern-5mu/\!}})e^{-2\kappa L}), 
\notag
\end{eqnarray}%
where $\kappa =\sqrt{\frac{\xi ^{2}}{c^{2}}+\left\vert \mathbf{k}_{%
\mathbin{\!/\mkern-5mu/\!}}\right\vert ^{2}}$, $R_{1}$$^{m}$ ($R_{2}$$^{m}$)
is the complex reflection coefficient of slab $1$ (slab $2$) in the $m$
polarization state ($s$ or $p$ states) and $k_{\mathbin{\!/\mkern-5mu/\!}}$
is the parallel component of the photon wave vector. The use of the complex
angular frequency $\omega $ $=$ $\mathit{i}\xi$ arises from numerical
computation considerations. Although this theory has been derived at zero temperature, it must be noticed that Eq.~\ref{U} can be perfectly used at room temperature provided that $\hslash \omega _{p}\gg k_{B}T$, a condition that is satisfied for all photon energies involved here as shown elsewhere \cite{s1,s2,s3,14,14c}. Such a condition means that the interaction reduces only to the effect of virtual photon exchange while the contribution of blackbody photons can be neglected. Independently of this consideration, the interaction energy given by Eq.~\ref{U} still exhibits a temperature dependence simply because the dielectric properties of materials -- permittivity values involved in the reflection coefficients -- are temperature dependent.

Let us take medium $1$ as the solid and medium $2$ as the
liquid. Hereafter, we propose a useful phenomenological
theoretical description of the superhydrophobicity tuning induced by the use
of a slab of metamaterial having ultra-broadband and
wide-angle absorption added on the flat interface of the medium $1$ in order to form a corrugated interface. For convenience, the reflection coefficient $R_{c}$ of the corrugated interface can be related to the reflection coefficient of the initially flat interface $%
R_{f}$, whatever the polarization state is, by using the ansatz:

\begin{equation}
R_{1,c}^{m}\rightarrow R_{1,f}^{m}\ \Lambda \left( \omega ,h, a\right),
\label{refc}
\end{equation}%
where $\Lambda \left( \omega ,h, a\right)$ is a function of the incident wave angular frequency $\omega$, of the metamaterial slab's geometry (layer thickness $h$) and its physical properties (effective absorption coefficient $a$, see below).

In the following, the metamaterial under interest is obtained from the flat interface by carving the surface of the solid across the depth $h$ between $0$ and $%
100 $ nm (or more), for instance, hence creating on the surface an array of
nanospikes -- roughly of conical or pyramidal shape -- with a typical base
about $10$ nm and a high aspect-ratio \cite{s1,s2}. The conical, or
pyramidal, shape of these nanospikes provides a layer with an effective
gradient index across the thickness $h$, acting as a antireflective layer
such that the corrugated interface now exhibits an ultra-broadband and
wide-angle absorption \cite{s1,s2}. This effect is well-known, for instance,
in black silicon where surface nanostructuring transforms a flat silicon
wafer into a nearly perfect black material \cite{bs1,bs2,bs3,bs4}. In this
case, it is well-known that \cite{bs1}:

\begin{equation}
\Lambda \left( \omega, h, a \right) =e^{-a(\omega /c)h},  \label{en}
\end{equation}%
where $a$ is an effective absorption coefficient here supposed to be constant against $%
\omega $. Obviously, this equation corresponds to the Beer--Lambert law with 
$a=2\mathtt{Im}\left\{ n\right\} =2n^{\prime \prime }$, which is applied to
an interface covered by an idealized perfect absorbing layer with a
thickness $h$ and with an effective optical index $n$ (see Appendix \ref{A} 
for the derivation of $n$). Theoretically, this model is justified provided that the reflection of the absorbing layer is indeed negligible, i.e. if the highly antireflecting
properties are allowed by an effective gradient index. Experimentally, this
simple model is very well justified for black silicon \cite{bs1,bs2,bs3,bs4}
for instance.

From Eqs. \ref{Hamaker},~\ref{U} and \ref{refc} it can be shown (see
Appendix \ref{B} for details) that the Hamaker constant can be recast as:

\begin{equation}
A_{H}=f\,A_{H,0},  \label{Hameff}
\end{equation}%
where $A_{H,0}$ is the Hamaker constant describing the interaction between the liquid
and the flat interface, i.e. without the metamaterial slab, and $f$ is a
function which describes the effect of the metamaterial. For a
metamaterial layer made of the array of cones described above \textit{via} Eq. \ref{en}%
, we obtain (see Appendix \ref{B}):%
\begin{equation}
f=\frac{1}{1+h/h_{0}}  \label{fcone}
\end{equation}%
It should be noted that, for an array of cylinders, $f$ is a constant which does not
depend on the metamaterial layer thickness $h$ \cite{s1,s3}. As justified
later herein, this makes cylinder-based metamaterials irrelevant to
demonstrate quantum effect on hydrophobicity \cite{s1,s3}.

The above derivation of the Hamaker constant is motivated by our will to compare theoretical predictions with experiments. It is well-known that Hamaker's theory is able to predict the equilibrium contact angle of liquid droplets on a surface \cite{vdW2}, in
general, and of water droplets in particular \cite{vdW4}. Indeed, from the
van der Waals potential energy calculated between a solid and a liquid, we
can immediately deduce \cite{vdW4} the corresponding contact angle $\theta $:

\begin{equation}
\cos (\theta )=-1+\frac{\left\vert U(L_{0})\right\vert }{\gamma _{l}},
\label{Wettrel}
\end{equation}%
where $U(L_{0})$ is the potential energy between the two media separated by
the distance $L_{0}$, which is the equilibrium separation distance \cite{14c}
between the water droplet and the surface. This distance, originally defined
for a flat surface \cite{14c}, remains about the same when the solid surface
is corrugated as shown elsewhere \cite{vdW4}. In Eq. \ref{Wettrel}, $\gamma _{l}$ is
the liquid surface tension.

As a result, from Eqs.~\ref{Hamaker},~\ref{Hameff}, and~\ref{Wettrel}, the
contact angle associated with the nanostructured surface can be easily
expressed through the relation:%
\begin{equation}
\cos (\theta )=-1+\left( 1+\cos (\theta _{0})\right) f
\label{WettingFormula}
\end{equation}%
which results from the quantum electrodynamics interpretation of the van der
Waals interfacial forces and where $\theta _{0}$ is the contact angle
on the flat surface. For the cone array slab, using Eq. \ref{fcone},
one gets: 
\begin{equation}
\cos (\theta )=-1+\left( 1+\cos (\theta _{0})\right) \frac{1}{1+h/h_{0}},
\label{WettingLaw}
\end{equation}%
By setting $h=0$ in Eq.~\ref{WettingLaw}, one recovers the contact angle
associated with the corresponding flat surface.

Nevertheless, it could be suggested that Eq.~\ref{WettingLaw} could be also
interpreted through the usual Cassie-Baxter model \cite{w1}, which
originates from a pure thermodynamic and geometrical analysis. We now
underline that such an alternative interpretation is not valid. Let us start
with the Cassie-Baxter expression \cite{w1}: 
\begin{equation}
\cos (\theta )=f_{1}\cos \theta _{1}+f_{2}\cos \theta _{2},  \label{CB}
\end{equation}%
with $f_{1}+f_{2}=1$, and where $f_{1}$ and $f_{2}$ are the fractional areas
of media composing the nanostructured surface, here respectively the air and
the substrate. Hence, $\theta _{1}$ would be the contact angle with the air,
i.e. $\theta _{1}=180^\circ $, and $\theta _{2}$ would be the contact angle
with the flat substrate, i.e. $\theta _{2}=\theta _{0}$. By assuming Eq.~\ref%
{CB} to be equivalent to Eq.~\ref{WettingLaw}, we would get for $f_{2}$ (and 
$f_{1}=1-f_{2}$):

\begin{equation}
f_{2}=\frac{1}{1+h/h_{0}}.  \label{f2}
\end{equation}

On the other hand, according to the Cassie-Baxter approach, $f_2$ is the
fractional area of liquid in contact with the surface $S_2$ of the substrate
material. By noting $S_1$ the area of the liquid/air interface, it comes:

\begin{equation}
f_{2}=\frac{S_{2}}{S_{1}+S_{2}}=\frac{1}{1+S_{1}/S_{2}}.  \label{f2b}
\end{equation}

Considering Eqs.~\ref{h0fin} and ~\ref{f2}, Eq.~\ref{f2b} could be
interpreted in the context of the Cassie-Baxter approach if and only if:

\begin{equation}
\frac{S_{1}}{S_{2}}=\frac{\pi h}{a_{0}}.  \label{ratio}
\end{equation}

However, even when neglecting air pressure and liquid weight, $S_{1}$ and $%
S_{2}$ must depend on the exact geometry of the nanostructure (spikes), on
the deformation of the liquid-air interface due to surface tension, and on
the extend of the wet surface at the top of the spikes. In our quantum
mechanically derived model, the ratio only depends on the height of the
spikes (not on their actual geometry) and on the lattice parameter of the
array of these spikes. As a matter of fact, there is no trivial geometrical
construction that allows us to accept such an identity (Eq.~\ref{ratio}),
except by pure coincidence. For this reason, the main outcome of our model,
i.e. Eq.~\ref{WettingLaw}, cannot be interpreted according to the
Cassie-Baxter model. From a fundamental point of view, Eq.~\ref{WettingLaw}
results from optically-controlled suppression of vacuum photon modes
responsible for van der Waals interaction. As a consequence, any
experimental evidences of a wetting behaviour following Eq.~\ref{WettingLaw}
would be a signature of a superhydrophobic effect induced by
optically-controlled suppression of vacuum photon modes. The next section
reports on our very first attempts to check this statement experimentally.

Before reporting and discussing results, we point out that, for a cylinder-based metamaterial slab, as $f$ is constant, it is always possible to relate $f$ to geometrical parameters, by resorting solely to a Cassie-Baxter model, for instance. For example, considering an hexagonal array of cylinders with radius $r_{0}$, one gets $f=(2\pi /\sqrt{3})(r_{0}/a_{0})^{2}$. Unfortunately, in this case and as shown elsewhere \cite{s1,s3}, 
\textit{ab initio} numerical computations based on the quantum derivation of
the van der Waals forces do not allow us to discriminate between quantum
contributions and thermodynamical ones in a Cassie-Baxter approach. This is
further discussed in the next section.

\section{Experimental results}

\begin{figure}[h!]
\centering
\includegraphics[width = \linewidth]{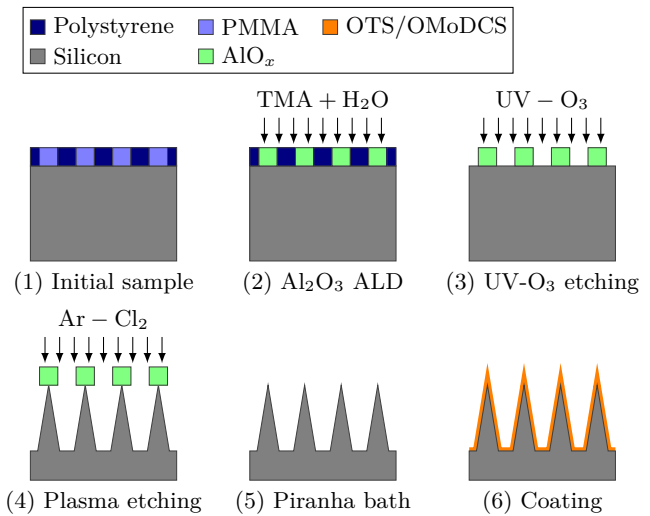}
\caption{(Color online). Sketch of each fabrication step 1 to 6 detailed in
the text. 1. Formation of honeycomb structure of PMMA self-assembled studs
in a PS matrix using block copolymer nano-manufacturing. 2. PMMA domain
conversion into Al$_x$O$_y$ (green) by sequential infiltration synthesis. 3.
Selective removal of the PS matrix using UV-O$_3$ treatment. 4. Silicon
etching by RIE with Al$_x$O$_y$ nanostructures as hard mask. 5. Hard mask
stripping. 6. Organosilane monolayer grafting.}
\label{fig2}
\end{figure}

\begin{table*}[ht!]
\centering
\renewcommand{\arraystretch}{1.5} 
\begin{tabular}{c|c|c|c|c|c}
Sample & $h$ (nm) & $a_{0}$ (nm) & $r_{0}$ (nm) & Graft & $\theta $ ($^\circ 
$) \\ \hline\hline
C35-5' & 200 $\pm$ 5 & 35 $\pm$ 2 & 15 $\pm$ 2 & OTS & 159 $\pm$ 3 \\ \hline
C35-3' & 120 $\pm$ 5 & 35 $\pm$ 2 & 15 $\pm$ 2 & OMoDCS & 130 $\pm$ 3 \\ 
\hline
C23-5' & 100 $\pm$ 5 & 23 $\pm$ 2 & 10 $\pm$ 2 & OMoDCS & 161 $\pm$ 3 \\ 
\hline
C23-3' & 85 $\pm$ 5 & 23 $\pm$ 2 & 8.5 $\pm$ 2 & OMoDCS & 142 $\pm$ 3%
\end{tabular}%
\caption{Samples used for wetting characterization (Cxx-y' with xx the
lattice parameter in nm and y the etching time in min). $h$: structure's
height, $a_0$: lattice parameter, $r_0$: structure's radius on surface,
Graft: Organosilane monolayer, $\protect\theta$: equilibrium contact angle.}
\label{t1}
\end{table*}

As an experimental platform to check our theoretical predictions,
nanopatterned silicon samples were fabricated using block copolymer
nano-manufacturing and coated with two kinds of organosilane monolayers:
octadecyltrichlorosilane (OTS) or octadecylmethoxydichlorosilane (OMoDCS).
These molecules bear the same aliphatic chain ($n=18$) and only differ due
to the substitution of a methoxy group (OMoDCS) instead of a chloride (OTS)
on the silicon atom. This substitution allows an easier control of the
surface grafting. Such a coating was motivated by our previous theoretical
works \cite{s1,s2} where nanopatterned surfaces of molecular materials --
such as polyethylene -- where considered. This avoided some difficulties to
construe van der Waals interactions with covalent materials. In the present
work, silicon facilitates manufacturing at the nanometer scale, but the
organic coating allows to keep the choice of molecular materials as a
working hypothesis\footnote{%
It can be proved that just few nanometers of coating material allows to
erase the effects of the covered medium \cite{s3,14c}.}.

The samples exhibit roughly hexagonal arrays of nanopillars or nanospikes.
In order to fabricate our samples, we followed a protocol inspired by the
work of Checco \textit{et al} \cite{checco,checco2} and we adapted it in
order to reach smaller lateral features (i.e. lattice parameter $a_0$) that
are mandatory for the problem under consideration. This constraint set a
huge challenge for experimentalists since we had to create spikes as close
as a few tens of nanometers to one another and as high as one hundred
nanometers.

We proceeded as follows (steps 1-3: formation of a nanopatterned mask on
silicon, steps 4-5: etching of silicon through the mask, steps 6: conformal
grafting of an organic monolayer on the nanostructured silicon surface, see
Fig.~\ref{fig2}):\newline

\begin{enumerate}
\item Silicon wafers were spin-coated with a poly(styrene)-b-poly(methyl
methacrylate) (PS-b-PMMA) block copolymer solution on a silicon substrate
grafted with a neutral statistical (PS-stat-PMMA) copolymer, and thermally
annealed, resulting in the formation of an honeycomb structure of PMMA
self-assembled studs in a PS matrix.

\item The PMMA core was substituted by aluminium oxide (Al$_x$O$_y$) by
sequential infiltration synthesis performed within an Atomic Layer
Deposition (ALD) apparatus using trimethylaluminum (TMA) and H$_2$O as
precursors.

\item The PS matrix was removed by UV-O$_3$ treatment.

\item Silicon was etched by RIE (Reactive Ion Etching with Ar and Cl$_2$
plasma in ratio 4:1) with the Al$_x$O$_y$ nanostructures as a hard mask.

\item The hard mask was removed in a bath of ``piranha'' solution (H$_2$O$_2$
+ H$_2$SO$_4$ in ratio 3:1).

\item Etched silicon surfaces were grafted with an organic layer of OTS or
OMoDCS (Degreasing: 10 min ultrasonication (US) in acetone + 10 min US in
methanol. Activation: 30 min in H$_2$SO$_4$/H$_2$O$_2$ (70:30) solution at
90 $^\circ $C. Grafting solutions: $4$ mM OTS in hexane (0 $^\circ $C, 60
min) or $2$ mM OMoDCS in hexane (RT, 120 min). Rinsing: 2 $\times$ 5 min US
in CHCl$_3$).
\end{enumerate}

Two different PS-b-PMMA block copolymers with different intrinsic
periodicity (i.e., C23 and C35 with cylinder-to-cylinder distance of 23 and
35 nm, respectively) were used to fabricate the corrugated patterns in order
to obtain different lattice parameters. In total, 17 samples have been
realized, but only 4 samples had the appropriate geometry for testing our
prediction after steps 2 to 5 (silicon etching and mask removal). This means
that the manufacturing process provides -- with a success rate of $24$$\%$
-- the samples with the expected lateral corrugation distance, with a weak
dispersion, and with patterns sufficiently similar to each other. These
samples were then grafted with organic molecules in order to emulate the
nanostructured molecular solid surface previously considered in our \textit{%
ab initio} theoretical model \cite{s1}. Hence, the initial nanostructured
silicon wafer served only as a mechanical support for the organic monolayer.

\begin{figure}[h!]
\centering
\includegraphics[width=8.5 cm]{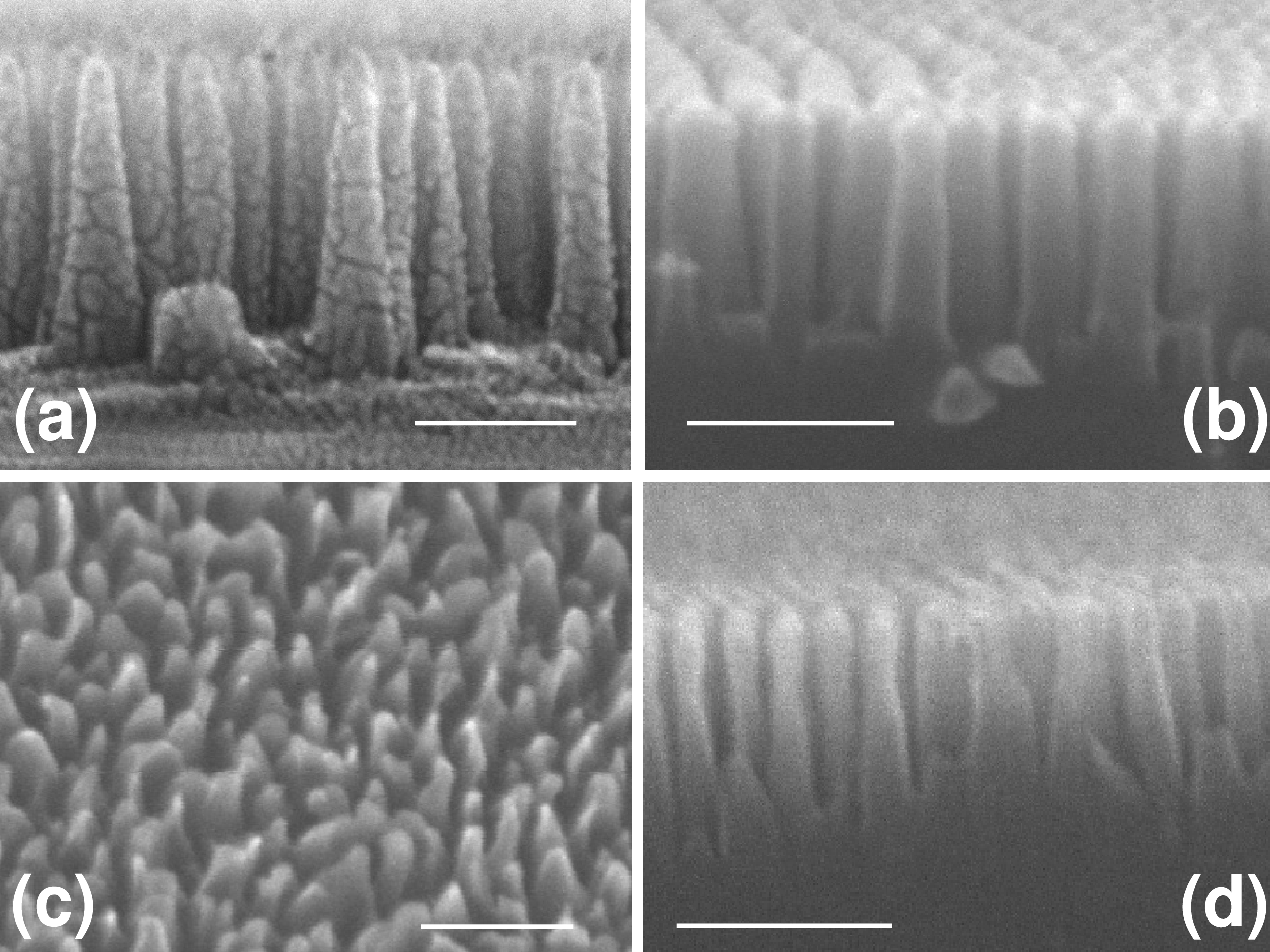}
\caption{Scanning electron microscopy images of nanopatterned silicon
surfaces for C35-5' sample (a), C35-3' sample (b), C23-5' sample (c) and
C23-3' sample (d). Scale bar: $100$ nm.}
\label{fig3}
\end{figure}

\begin{figure}[ht!]
\centering
\includegraphics[width = .95\linewidth]{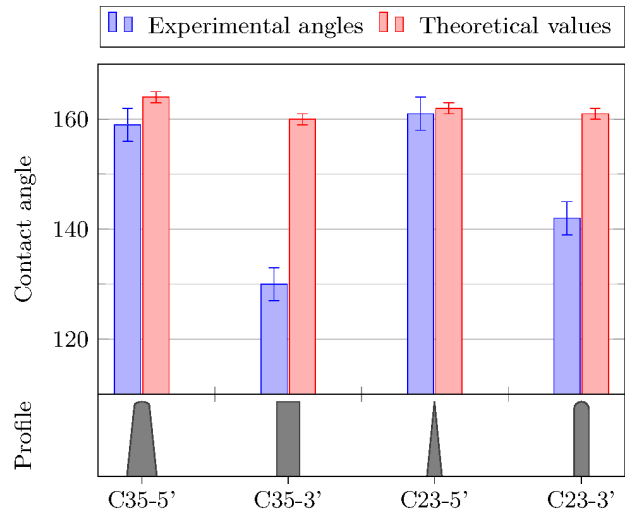}
\caption{(Color online). Comparison for each kind of patterns between
experimental (blue) contact angles (in degrees) and theoretical ones (red)
from Eq.~\protect\ref{WettingLaw} assuming quantum contributions inducing
superhydrophobicity. The patterns are classified according to two extreme
shapes, i.e. pillars (cylinders) and spikes (sharp cones).}
\label{fig4}
\end{figure}

The properties of these samples are summarized in Table 1 and their typical
surface pattern are shown in Fig.~\ref{fig3} by scanning electron
microscopy (SEM). Equilibrium contact angle measurements were performed on these
samples and the corresponding results are reported in Table 1. The measured
angle values and their uncertainties are also shown in Fig.~\ref{fig4} (blue
bars). For the flat surfaces, organosilane layer thickness determined by
spectroscopic ellipsometry after optimisation of the grafting process were
in agreement with the literature values (OTS: $3.0 \pm 0.8$ nm and OMoDCS: $%
2.3 \pm 0.3$ nm). For the same surface, the advancing contact angles were $%
109 \pm 1 ^\circ $ (OTS) and $110 \pm 1 ^\circ $ (OMoDCS). Nanostructuration
clearly induces an increase of the advancing contact angles. On the basis of
usually accepted criteria for superhydrophobicity (contact angle larger than 
$150^\circ $ and low contact angle hysteresis), samples C35-5' and C23-5',
respectively corresponding to the cones and spikes nanostructures, can be
considered as superhydrophobic.

Using Eq.~\ref{WettingLaw}, we computed the theoretical contact angles using
the reported experimental flat surface contact angle, as well as the height $%
h$ and the periodicity $a_{0}$ of the nanostructures which were estimated
from SEM observations. The observed nanostructures were classified according
to two extreme shapes, namely cylindrical pillars and sharp cones (spikes).
The results are reported in Fig.~\ref{fig4} (red bars) together with
uncertainties resulting from error propagation of the experimental inputs.
As expected from our theoretical analysis, since pillar-like nanostructures
offer no significant anti-reflecting properties in UV domain \cite{s1,s2,s3}
(due to the absence of effective index gradient), Eq.~\ref{WettingLaw}
cannot be applied to C35-3' and C23-3' nanostructures, which explains the
differences between theoretical and measured angles. By contrast, Eq.~\ref%
{WettingLaw} should be highly relevant to spikes (samples C35-5' and C23-5')
since they allow for the graded-index profile at the origin of the
optically-controlled suppression of vacuum photon modes. Indeed, one can
observe that experimental and theoretical contact angles match very well in
both C35-5' an C23-5' samples (see Fig.~\ref{fig4}). However, it is worth
noticing that general and complex exact numerical computations using Eq.~\ref%
{U} for cylindrical pillars -- previously achieved for polyethylene \cite{s3}
-- led to a contact angle about $140^{\circ }$ similar to those of samples
C35-3' and C23-3', thus strengthening our approach derived from Eq.~\ref{U}.
Nevertheless, as explained in the previous section, the theoretical
behaviour of the hydrophobic properties of the cylindrical-based effective
layer given by Eq. \ref{WettingFormula} ($f$ being constant against $h$) can
be also discussed, for instance, on the basis of the Cassie-Baxter approach (with $%
f=(2\pi /\sqrt{3})(r_{0}/a_{0})^{2}$). Then, using values of $r_{0}$ and $%
a_{0}$ given in Table 1, Eq. \ref{WettingFormula} leads to the following
contact angles: $124\pm 9^{\circ }$ (C35-3') and $132\pm 13^{\circ }$
(C23-3'). Despite large uncertainties, these predicted values are compatible with the
measured contact angles. In conclusion, not only the cylindrical pillars do not
enable to obtain superhydrophobicity, but they do not allow us to discriminate
between classical effects and quantum ones.

It could be then suggested that the present results -- regarding surfaces covered with cones and spikes -- could be also interpreted through a classical Cassie-Baxter model. For instance, instead of perfect cones, and as it could be suggested by SEM images, one could consider truncated cones surmounted by hemispheres with radius $r_{top}$. In such a case, the contact angle is given by the well-known relation \cite{28a}:
\begin{equation}
\cos {\theta}=n\pi r_{top}^2(1+\cos {\theta }_0)^2-1,  \label{CB2}
\end{equation}
where $n$ is the number of hemispheres per unit area for a hexagonal array, i.e. $n=2/(a_0^2\sqrt 3)$. However, fitting $r_{top}$ in order to match experimental data leads then to hemispheres' radii two times larger than those expected from SEM measurements. This discrepancy clearly invalidates such a Cassie-Baxter approach.
It could also be objected that more complex classical models, for instance using
finite element methods relying on Navier-Stokes equations \cite{w2,w3,w4}%
, could maybe explain also these results. Nevertheless, the experimental
study of wetting models on corrugated surfaces with lateral corrugations as
close as $10$ nm is complicated by complex manufacturing processes -- hard
to replicate -- and by huge uncertainties on contact angle measurements \cite%
{un1,un2}. This leaves us with a narrow experimental window in order to
definitively validate or invalidate our model. Moreover, as shown by Eq. \ref{CB2} for instance, superhydrophobic behaviour mainly depends on the geometrical properties of the top of the nanostructures in the Cassie-Baxter approach, while the quantum description given by Eq.~\ref{WettingLaw} underlines the importance of the nanostructures' height.
As a consequence, any
remaining doubts on the exact mechanisms at play will only be resolved if
one is able to improve the reliable production of a large number of samples
with different profiles in order to increase the significance of the present experimental results and to be able to compare the classical and quantum models, for instance by looking for a dependance of the contact angle on the nanostructures' height.
We hope that the present study will stimulate further research in this direction.

\section{Conclusions and perspectives}

Previous theoretical works suggested that, superhydrophobicity can be
induced through the use of a nanostructured surface that is designed to form
a thin metamaterial layer with ultra-broadband and wide-angle absorption.
This layer precludes the exchange of virtual photons and induces the
collapse of the van der Waals force allowing to reach superhydrophobicity.
We have given an interpretation of this fundamental concept through a
phenomenological approach which allowed us to derive a simple effective
contact angle formula that is the typical signature induced by quantum
effects on superhydrophobicity. Using advanced masking and etching
techniques for silicon wafer texturing at the nanometer scale and subsequent
grafting of organic monolayers, we have realized unique deeply
nanostructured surfaces covered by arrays of nanospikes or nanopillars in
order to provide a plateform for assessing our model. In samples exhibiting
nanospikes, we have measured static contact angles which could be
interpreted as a result of the suppression of quantum fluctuations as
predicted by our model. Further similar experiments will be considered in a
next work in order to reinforce the first assessment of the model provided
by these preliminary results.

\section*{ACKNOWLEDGMENTS}

The authors thank Alexandre Felten and Francesca Cecchet for useful
discussions and reading of the manuscript. This research used resources of the Electron Microscopy Service located at the University of Namur. This Service is member of the ''Plateforme Technologique Morphologie -- Imagerie'' (MORPH-IM). G.F. gratefully acknowledges
Arkema for providing the materials to prepare the block copolymer mask.

\section*{DATA AVAILABILITY}

The data that support the findings of this study are available from the
corresponding authors upon reasonable request.

\appendix

\section{Metamaterial effective optical index}

\label{A}

When the surface is corrugated by an array of conical pillars, two effects occur: diffraction modes arising from the periodic lateral
corrugation and refractive index gradient arising from the vertically
aligned cones.

\begin{figure}[ht]
\centering
\includegraphics[width = \linewidth]{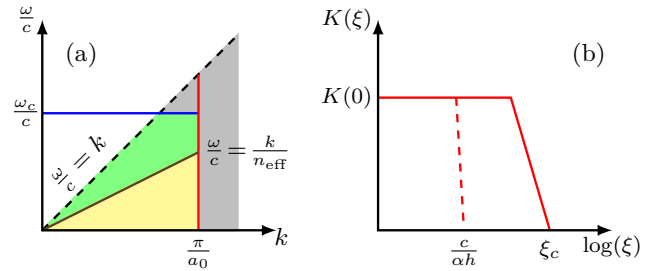}
\caption{(Color online). (a) Virtual photon dispersion between interfaces
(grey) with frequency cut-off (permittivity cut-off) and wavevector cut-off
(due to periodic array). Green and orange areas are equal and define the
average dispersion relation and then the effective index $n_{\text{eff}}$.
(b) The $K(\protect\xi)$ function, Eq.~\protect\ref{K}, presents a typical
low-band pass behaviour with a cut-off frequency at $\protect\xi_c$, which
is illustrated here. The dashed line illustrates the cut-off that is
introduced in the Hamaker constant expression, Eq.~\protect\ref{HamForm}, by
the factor $e^{-\protect\alpha (\protect\xi /c)h}$.}
\label{fig1}
\end{figure}

First, in the long wavelength limit, a grating effective index $n_{\text{eff}%
}$ can be defined, which is related to the wavevector cut-off effect due to
the lateral corrugation characterized through the period $a_{0}$. Let us now
estimate $n_{\text{eff}}$. As the medium is constituted by a set of periodic
nanospikes with a period $a_{0}$, and since one deals with wavelengths
greater than $a_{0}$, all diffraction orders of the periodically structured
surface (i.e. grating) are evanescent except for the zeroth-order of
diffraction (specular reflection). As a result, the electromagnetic field
can efficiently penetrate the medium while being exponentially damped,
allowing for a strong effective absorption of light. Since the surface has
discrete translational symmetry in lateral directions (cf. periodical
array), $k_{\mathbin{\!/\mkern-5mu/\!}}$ should present a cut-off typically
near $\pi /a_{0}$ at the border of the Brillouin zone. In addition, $\omega $%
, and thus $\xi $, should present a typical cut-off $\xi _{c}\sim \omega
_{c} $ beyond which the permittivity of the material tends to the vacuum
permittivity. This consideration is general and valid for any material, as
the frequency tends to infinity. In practice $\omega _{c}$ can be considered
as finite. In addition, for short distances between interacting bodies,
evanescent waves ($\frac{\omega ^{2}}{c^{2}}n_{\text{eff}}^{2}-\left\vert 
\mathbf{k}_{\mathbin{\!/\mkern-5mu/\!}}\right\vert ^{2}<0$) dominate and
must follow the dispersion relation $\frac{\omega }{c}<k_{%
\mathbin{\!/\mkern-5mu/\!}}/n_{\text{eff}}$. As a result, possible virtual
photons parameters $(\omega ,k_{\mathbin{\!/\mkern-5mu/\!}})$ occupy the
domains in $(\omega ,k_{\mathbin{\!/\mkern-5mu/\!}})$-space that are
depicted in green and orange in Fig.~\ref{fig1}a. Considering both
wavevector and frequency finite limits, we decide to describe $n_{\text{eff}%
} $ as the effective index which corresponds to the average dispersion
relation separating green and orange domains into two equal parts. Then,
from geometrical considerations, one easily shows that:%
\begin{equation}
n_{\text{eff}}=\frac{2x^{2}}{2x-1}\text{ with }x=\frac{\pi c}{a_{0}\xi _{c}}.
\label{neff}
\end{equation}%
In the following, we will assume that $\xi _{c}$ is small enough compared to 
$\pi c/a_{0}$ so that\footnote{$a_{0}\approx 10$ nm, i.e. $\pi
c/a_{0}\approx 10^{17}$ rad$\cdot $s$^{-1}$, while typical permittivity
cut-off is about \cite{25} $10^{16}$ rad$\cdot $s$^{-1}$. Therefore $x$ is
of the order of $10$ and the assumption $x\gg 1$ made to approximate $n_{%
\text{eff}}$ by $x$ from Eq.~\ref{neff} is fully justified.} $n_{\text{eff}%
}\sim x$. In practice, $x$ is of the order of ten.

Next, always in the long wavelength limit, the profile of the corrugation
can lead to a refractive index gradient \cite{s1,s2} $n_{g}(z)$ while $z$
varies between the substrate $(z=0)$ and the top of the corrugation $(z=h)$
in vacuum. For instance -- for a cone array along an hexagonal lattice --
one gets \cite{s1}: 
\begin{equation}
n_{g}(z)=\sqrt{1+\gamma (n_{\text{eff}}^{2}-1)(z-h)^{2}/h^{2}},  \label{ng}
\end{equation}%
where $\gamma =\pi /(2\sqrt{3})$ is the 2D filling rate of the close-packed
hexagonal lattice at the bottom of the cone ($z=0$). For cylindrical
pillars, on the other hand, it is worth noting that such a gradient does not exist but also that Eq.~\ref{en} cannot be applied as explained above  (end of section \ref{A}). In the present phenomenological approach, we replace $n_{g}(z)$ by a spatially averaged
index which must correspond to the above mentioned effective optical index $%
n $ of the absorbing layer. To derive such a mean index, one considers the
optical path $\mathcal{L}$ through the absorbing layer of thickness $h$ such
as:

\begin{equation}
\mathcal{L=}\int_{0}^{h}n_{g}(z)\mathrm{d}z=nh,  \label{path}
\end{equation}

with%
\begin{equation}
n=\frac{1}{h}\int_{0}^{h}n_{g}(z)\mathrm{d}z.  \label{mean}
\end{equation}

It can be shown that\footnote{%
Using the expression of Eq.~\ref{ng} in Eq.~\ref{mean}, and setting $u=1-z/h$
and $p^{2}=\gamma (n_{\text{eff}}^{2}-1)$, one gets $n=\int_{0}^{1}\sqrt{%
1+p^{2}u^{2}}\mathrm{d}u=\frac{1}{2}\left( \sqrt{1+p^{2}}+\arg \sinh
(p)/p\right) $. For $p$ great enough, $n\sim p/2$, and thus, for $n_{\text{%
eff}}$ great enough -- in the present context $n_{\text{eff}}\approx 5$
fulfilling the condition -- $n\sim (\sqrt{\gamma }/2)n_{\text{eff}}\approx
(1/2)\,n_{\text{eff}}$.} 
\begin{equation}
n\sim (1/2)\,n_{\text{eff}}.  \label{eff}
\end{equation}

Considering Eq.~\ref{eff}, we are able to find the searched expression for $%
\alpha $:%
\begin{eqnarray}
\alpha &=&2n\sim n_{\text{eff}}\sim x  \notag \\
&\sim &\frac{\pi c}{a_{0}\xi _{c}}.  \label{alpha}
\end{eqnarray}

\section{Hamaker constant derivation}

\label{B}

In the case where the surface of body $1$ is corrugated the expression of the potential energy, Eq. \ref{U}, becomes, thanks to Eq. \ref{refc} (the subscripts on the reflection coefficients in Eq. \ref{refc} are dropped for conciseness):
\begin{eqnarray}
U(L) &=&\frac{\hbar }{2\pi }\sum_{m=s,p}\int \frac{\mathrm{d}^{2}k_{%
\mathbin{\!/\mkern-5mu/\!}}}{(2\pi )^{2}}\int_{0}^{\infty }\mathrm{d}\xi
\label{U1} \\
&&\times \ln (1-R_{1}^{m}(i\xi ,\mathbf{k}_{\mathbin{\!/\mkern-5mu/\!}%
})\Lambda \left( i\xi ,h, \alpha\right) R_{2}^{m}(i\xi ,\mathbf{k}_{%
\mathbin{\!/\mkern-5mu/\!}})e^{-2\kappa L}),  \notag
\end{eqnarray}%
Considering Eq.~\ref{en} where $a$ is now substituted by\footnote{%
Indeed, as $\exp (-a_{\omega }(\omega /c)h)=\exp (-2(\omega /c)\text{Im}%
\left\{ n(\omega )\right\} h)=\text{Re}\left\{ \exp (i2(\omega /c)n(\omega
)h)\right\} $, then $\exp (-a_{i\xi }(i\xi /c)h)=\text{Re}\left\{ \exp
(i2(i\xi /c)n(i\xi )h)\right\} =\exp (-2(\xi /c)n(i\xi )h)$ since $n(i\xi )$
is real.} $\alpha =2n(i\xi )$, where we further assume that $n(i\xi )\sim n$
is constant, i.e. frequency-independent on the domain of interest. A tricky
part of the present phenomenological approach lies in the estimation of the
effective optical index $n$ of the metamaterial layer, which is detailed in Appendix \ref{A}.

Let us first rewrite Eq.~\ref{U1} as:%
\begin{eqnarray}
U(L) &=&\frac{\hbar }{4\pi ^{2}L^{2}}\sum_{m=s,p}\int_{0}^{\infty }q\mathrm{d%
}q\int_{0}^{\infty }\mathrm{d}\xi  \label{U2} \\
&&\times \ln (1-R_{1}^{m}(i\xi ,q/L)\Lambda \left( i\xi ,h, \alpha\right)
R_{2}^{m}(i\xi ,q/L)e^{-2\rho }),  \notag
\end{eqnarray}%
where we used: $(1/(2\pi )^{2})\int \mathrm{d}^{2}k_{\mathbin{\!/\mkern-5mu/\!}%
}=(1/2\pi )\int k_{\mathbin{\!/\mkern-5mu/\!}}\mathrm{d}k_{%
\mathbin{\!/\mkern-5mu/\!}}$, and $\rho =\sqrt{\frac{\xi ^{2}}{c^{2}}%
L^{2}+q^{2}}$, with $q=k_{\mathbin{\!/\mkern-5mu/\!}}L$. Using Eq.~\ref%
{Hamaker}, the effective Hamaker constant $A_{H}$ is directly
deduced from Eq.~\ref{U2}:%
\begin{eqnarray}
A_{H} &=&-\frac{3\hbar }{\pi }\sum_{m=s,p}\int_{0}^{\infty }q\mathrm{d}%
q\int_{0}^{\infty }\mathrm{d}\xi  \label{Ham} \\
&&\times \ln (1-R_{1}^{m}(i\xi ,q/L)\Lambda \left( i\xi ,h, \alpha\right)
R_{2}^{m}(i\xi ,q/L)e^{-2\rho }).  \notag
\end{eqnarray}%
\bigskip

Recalling the expressions of the Fresnel coefficients for flat interfaces:%
\begin{eqnarray}
R_{1(2)}^{s}(\omega ,k_{\mathbin{\!/\mkern-5mu/\!}}) &=&\frac{%
k_{z,3}-k_{z,1(2)}}{k_{z,3}+k_{z,1(2)}},  \label{Rs} \\
R_{1(2)}^{p}(\omega ,k_{\mathbin{\!/\mkern-5mu/\!}}) &=&\frac{\varepsilon
_{3}k_{z,1(2)}-k_{z,3}\varepsilon _{1(2)}}{\varepsilon
_{3}k_{z,1(2)}+k_{z,3}\varepsilon _{1(2)}},  \label{Rp}
\end{eqnarray}%
where $k_{z,i}=\sqrt{\frac{\omega ^{2}}{c^{2}}\varepsilon _{i}-k_{%
\mathbin{\!/\mkern-5mu/\!}}^{2}}$, we get, since $k_{\mathbin{\!/\mkern-5mu/%
\!}}=q/L$: 
\begin{equation}
\lim_{L\rightarrow 0}R_{1(2)}^{s}(i\xi ,q/L)=0,  \label{limRs}
\end{equation}%
and%
\begin{equation}
\lim_{L\rightarrow 0}R_{1(2)}^{p}(i\xi ,q/L)=\frac{\varepsilon _{3}(i\xi
)-\varepsilon _{1(2)}(i\xi )}{\varepsilon _{3}(i\xi )+\varepsilon
_{1(2)}(i\xi )}.  \label{limRp}
\end{equation}%
Then, in the limit where $L$ tends to zero: 
\begin{eqnarray}
A_{H} &=&-\frac{3\hbar }{\pi }\int_{0}^{\infty }q\mathrm{d}q\int_{0}^{\infty
}\mathrm{d}\xi  \label{approx1} \\
&&\times \ln (1-K(\xi )e^{-2q}\Lambda \left( i\xi ,h, \alpha\right) ),  \notag
\end{eqnarray}%
with 
\begin{equation}
K(\xi )=\frac{\varepsilon _{3}(i\xi )-\varepsilon _{1}(i\xi )}{\varepsilon
_{3}(i\xi )+\varepsilon _{1}(i\xi )}\frac{\varepsilon _{3}(i\xi
)-\varepsilon _{2}(i\xi )}{\varepsilon _{3}(i\xi )+\varepsilon _{2}(i\xi )}.
\label{K}
\end{equation}%
For most of usual materials, it can be verified that $K(\xi )\ $is small
enough such that $\ln (1-K(\xi )e^{-2q}\Lambda \left( i\xi ,h, \alpha\right) )\sim
-K(\xi )e^{-2q}\Lambda \left( i\xi ,h, \alpha\right) $. In that case, the integral
over $q$ can be solved analytically, so that Eq.~\ref{approx1} becomes: 
\begin{eqnarray}
A_{H} &\sim &\frac{3\hbar }{\pi }\int_{0}^{\infty }q\mathrm{d}%
q\int_{0}^{\infty }\mathrm{d}\xi K(\xi )e^{-2q}\Lambda \left( i\xi ,h, \alpha\right)
\notag \\
&=&\frac{3\hbar }{4\pi }\int_{0}^{\infty }\mathrm{d}\xi K(\xi )\Lambda
\left( i\xi ,h, \alpha\right) .  \label{HamForm}
\end{eqnarray}%
For a flat interface (i.e. $h=0$, for instance), Eq.~\ref{HamForm} reduces
to:%
\begin{eqnarray}
A_{H,0} &=&\frac{3\hbar }{4\pi }\int_{0}^{\infty }\mathrm{d}\xi K(\xi ) 
\notag \\
&=&\frac{3\hbar }{4\pi }\int_{0}^{\infty }\mathrm{d}\xi  \label{HamTh0} \\
&&\times \frac{\varepsilon _{3}(i\xi )-\varepsilon _{1}(i\xi )}{\varepsilon
_{3}(i\xi )+\varepsilon _{1}(i\xi )}\frac{\varepsilon _{3}(i\xi
)-\varepsilon _{2}(i\xi )}{\varepsilon _{3}(i\xi )+\varepsilon _{2}(i\xi )},
\notag
\end{eqnarray}%
which is the well-known expression of the Hamaker constant at absolute zero
temperature albeit valid up to the room temperature \cite{s3,14c}.

On the other hand, using Lorentz description \cite{25} of the dielectric
functions $\varepsilon (i\xi )$ and using the expression for $K(\xi )$ (see
Eq.~\ref{K}), one can verify that $K(\xi )$ possesses globally a low-pass
spectral behaviour shown in Fig.~\ref{fig1}b. As $K(\xi )$ is almost
constant until the cut-off is reached at $\xi _{c}$, one can approximate $%
A_{H,0}$ by: 
\begin{eqnarray}
A_{H,0} &\sim &\frac{3\hbar }{4\pi }K(0)\xi _{c}  \label{approxA0} \\
&\sim &\frac{3\hbar }{4\pi }\frac{\varepsilon _{3}(0)-\varepsilon _{1}(0)}{%
\varepsilon _{3}(0)+\varepsilon _{1}(0)}\frac{\varepsilon
_{3}(0)-\varepsilon _{2}(0)}{\varepsilon _{3}(0)+\varepsilon _{2}(0)}\xi
_{c}.  \notag
\end{eqnarray}%
Now, in Eq.~\ref{HamForm}, that is for a nanostructured interface, the
factor $\Lambda \left( i\xi ,h, \alpha\right) = e^{-\alpha (\xi /c)h}$ (from Eq.~\ref{en}) introduces a new cut-off frequency at $%
c/\alpha h$, which depends on $h$. A careful analysis shows that\footnote{%
As a heuristic argument, as $K(\xi )$ is almost constant until its cut-off
at $\xi _{c}$, we can write:\newline
$A_{H}=\frac{3\hbar }{4\pi }\int_{0}^{\infty }\mathrm{d}\xi K(\xi
)e^{-\alpha (\xi /c)h}\newline
\sim \frac{3\hbar }{4\pi }\int_{0}^{\infty }\mathrm{d}\xi K(0)e^{-\xi /\xi
_{c}}e^{-\alpha (\xi /c)h}\newline
\sim \frac{3\hbar }{4\pi }K(0)\int_{0}^{\infty }\mathrm{d}\xi e^{-\xi \left(
1/\xi _{c}+\alpha h/c\right) }$,\newline
since the integral is essentially determined by the cut-off frequency
imposed by $\alpha $ (see Fig.~\ref{fig1}b) and therefore $K$ can be
considered as constant in the integral. Now, if $h=0$, we have: $A_{H,0}=%
\frac{3\hbar }{4\pi }\int_{0}^{\infty }\mathrm{d}\xi K(\xi )e^{-\xi /\xi
_{c}}\sim \frac{3\hbar }{4\pi }\xi _{c}K(0)$. As a result: $A_{H}=\frac{%
3\hbar }{4\pi }\xi _{c}K(0)/(1+\alpha \xi _{c}h/c)=A_{H,0}/(1+\alpha \xi
_{c}h/c)$.}:

\begin{equation}
A_{H}\sim A_{H,0}\frac{1}{1+h/h_{0}},  \label{ExpAfin}
\end{equation}%
with $h_{0}=c/(\xi _{c}\alpha )$. Using the approximation of Eq.~\ref{alpha}
which relates $\alpha $ to $a_{0}$ and $\xi _{c}$, we get:%
\begin{equation}
h_{0}=\frac{a_{0}}{\pi }.  \label{h0fin}
\end{equation}%
Then, Eqs.~\ref{ExpAfin} is the searched expression, Eq. \ref{Ahfit},  
giving the dependence of Hamaker constant against the cone height $h$, 
\textit{i.e.} the thickness $h$ of the effective broadband absorber metamaterial
under consideration.
\\

\end{document}